\documentclass[twocolumn,prx,tightenlines,superscriptaddress,showpacs]{revtex4-1}

\usepackage{amsmath}
\usepackage{amssymb,amsfonts,latexsym}
\usepackage{bm}
\usepackage[mathcal]{euscript}
\usepackage{graphicx}
\usepackage{epsfig}
\usepackage{color}
\usepackage{xfrac}

\begin{document}

\title{Active Matter Class with Second-Order Transition to Quasi-Long-Range Polar Order}

\author{B. Mahault}
\affiliation{Service de Physique de l'Etat Condens\'e, CEA, CNRS, Universit\'e Paris-Saclay, CEA-Saclay, 91191 Gif-sur-Yvette, France}

\author{X.-c. Jiang}
\affiliation{Center for Soft Condensed Matter Physics and Interdisciplinary Research, Soochow University, Suzhou 215006, China}

\author{E. Bertin}
\affiliation{LIPHY, Universit\'e Grenoble Alpes and CNRS, F-38000 Grenoble, France}

\author{Y.-q. Ma}
\affiliation{National Laboratory of Solid State Microstructures and Department of Physics, Nanjing University, Nanjing 210093, China}
\affiliation{Center for Soft Condensed Matter Physics and Interdisciplinary Research, Soochow University, Suzhou 215006, China}

\author{A. Patelli}
\affiliation{Service de Physique de l'Etat Condens\'e, CEA, CNRS, Universit\'e Paris-Saclay, CEA-Saclay, 91191 Gif-sur-Yvette, France}

\author{X.-q. Shi}
\affiliation{Center for Soft Condensed Matter Physics and Interdisciplinary Research, Soochow University, Suzhou 215006, China}
\affiliation{Service de Physique de l'Etat Condens\'e, CEA, CNRS, Universit\'e Paris-Saclay, CEA-Saclay, 91191 Gif-sur-Yvette, France}

\author{H. Chat\'{e}}
\affiliation{Service de Physique de l'Etat Condens\'e, CEA, CNRS, Universit\'e Paris-Saclay, CEA-Saclay, 91191 Gif-sur-Yvette, France}
\affiliation{Beijing Computational Science Research Center, Beijing 100094, China}
\affiliation{Sorbonne Universit\'e, CNRS, Laboratoire de Physique Th\'eorique de la
Mati\`ere Condens\'ee, 75005 Paris, France}

\begin{abstract}
We introduce and study in two dimensions a new class of dry, aligning, active matter that exhibits a direct transition to orientational order,
without the phase-separation phenomenology usually observed in this context.
Characterized by self-propelled particles with velocity reversals and
ferromagnetic alignment of polarities, systems in
this class display quasi-long-range polar order with continuously-varying scaling exponents and yet
a numerical study of the transition leads to conclude that it does {\it not} belong to the
Berezinskii-Kosterlitz-Thouless universality class, but is best described as a standard critical point with algebraic divergence of correlations.
We rationalize these findings by showing that the interplay between order and density changes the role of defects.
\end{abstract}

\maketitle

Dry active matter refers to systems of mobile agents for which the surrounding fluid can be neglected
\cite{ramaswamy2010mechanics,marchetti2013hydrodynamics}.
A wealth of spectacular phenomena has been uncovered, in particular in situations where some local alignment of velocities arises,
such as animal groups \cite{cavagna2010scale,vicsek2012collective}, bacteria and cells crawling on a substrate \cite{zhang2010collective,peruani2012collective,nishiguchi2017long}, motility assays \cite{sumino2012large}, or vertically-shaken granular particles \cite{narayan2007long,deseigne2010collective,kumar2013flocking,weber2013long}.

We now have a satisfactory theoretical understanding of dry, aligning, active matter in the dilute limit, thanks to a
series of works at particle, hydrodynamic, and kinetic levels \cite{vicsek1995novel,toner1995long,toner1998flocks,ramaswamy2003active,gregoire2004onset,toner2005hydrodynamics,bertin2006boltzmann,chate2006simple,baskaran2008enhanced,chate2008collective,bertin2009hydrodynamic,baskaran2010nonequilibrium,ginelli2010large,mishra2010dynamic,ihle2011kinetic,farrell2012pattern,grossmann2012active,peshkov2012nonlinear,romanczuk2012mean,toner2012reanalysis,bertin2013mesoscopic,solon2013revisiting,caussin2014emergent,ngo2014large,peshkov2014boltzmann,putzig2014phase,thuroff2014numerical,solon2015phase,grossmann2016mesoscale,shankar2017low}.
In these systems, the dominating interaction, local alignment, is in competition with noise, and can lead to orientationally-ordered phases
that are endowed with generic long-range correlations and anomalous fluctuations \cite{toner1995long,toner1998flocks,ramaswamy2003active,chate2006simple,chate2008collective,ginelli2010large,mishra2010dynamic,ngo2014large}.
The emergence of order is {\it not} a direct, continuous phase transition, but occurs via phase-separation between a disordered gas and an
ordered liquid separated by a coexistence phase \cite{solon2013revisiting,putzig2014phase,solon2015phase}.
Genuine critical behavior has only been found when long-range interactions are present, brought by non-metric, ``topological" neighbors \cite{ginelli2010relevance,peshkov2012continuous} or by imposing
some incompressibility condition \cite{chen2015critical}.

Vicsek-style models,
which consist of constant-speed point particles that locally align their velocities in competition with some noise,
 have been instrumental in this success.
Their simplicity allows for both in-depth numerical study and the controlled derivation of hydrodynamic theories
\cite{bertin2006boltzmann,baskaran2008enhanced,bertin2009hydrodynamic,ihle2011kinetic,farrell2012pattern,grossmann2012active,romanczuk2012mean,peshkov2012nonlinear,peshkov2014boltzmann}.
The main classes of dry aligning active matter studied so far each possess a Vicsek-style representative.
Polar particles with ferromagnetic alignment --- the case of the original Vicsek model \cite{vicsek1995novel}---
give rise to true long-range polar order \cite{toner1995long,toner1998flocks,toner2012reanalysis} and a coexistence phase made of quantized traveling bands \cite{gregoire2004onset,chate2008collective,thuroff2014numerical,caussin2014emergent,solon2015pattern,solon2015phase}.
Nematic alignment leads to global nematic order and a chaotic coexistence phase mediated by unstable nematic bands \cite{chate2006simple,bertin2013mesoscopic,ngo2014large,grossmann2016mesoscale}.
For polar particles nematic order seems to be long-ranged \cite{ginelli2010large,shankar2017low,nishiguchi2017long}, whereas for finite velociy-reversal rate it is only quasi-long-ranged.

In this Letter we show that self-propelled particles with velocity reversals and local ferromagnetic alignment
exhibit novel collective properties and in particular a continuous transition to order.
In the restricted Vicsek setting where a particle's polarity is simply given by its velocity, no order can emerge in this case.
Here we relax this ``Vicsek constraint" by conferring particles a polarity that they align with that of neighbors while they move either along or against it.
Using kinetic and hydrodynamic-level descriptions derived from the microscopic model, we show that the analytic structure of this problem
is qualitatively different from that of the three other classes. In particular,
it is deprived from the generic linear instability at the root of the liquid/gas phase separation scenario.
Particle-level simulations confirm this: the emerging polar order is only quasi-long-ranged with continuously-varying scaling exponents,
while showing giant number fluctuations.
Thus, this case possesses many of the properties of the (equilibrium) $XY$ model.
Yet, surprisingly, a numerical study of the continuous ordering transition leads to conclude that it does {\it not} belong to the
Berezinskii-Kosterlitz-Thouless universality class \cite{berezinskii1971destruction,kosterlitz1973ordering,kosterlitz1974critical}
characteristic of the $XY$ model, but is best described as a standard critical point with algebraic divergence of correlations.
We rationalize these findings by showing that the coupling between order and density
deprives defects from their usual role.

We first define our ``Vicsek-shake" model. We restrict ourselves to
two space dimensions and square domains of linear size $L$ with periodic boundary conditions.
The velocity ${\bf v}_i$ of particle $i$ is given by ${\bf v}_i = \pm v_0 {\bf p}_i$ where the unit vector
${\bf p}_i$ is the particle's intrinsic polarity, and the sign is changed with probability $\alpha$ at each unit timestep.
Positions and polarities are governed by:
\begin{eqnarray}
{\bf r}_i(t+1) & = & {\bf r}_i(t) + {\bf v}_i(t+1) \label{model_r}\\
{\bf p}_i(t+1) & = & \left( {\cal R}_\eta \circ \Pi  \right) \langle {\bf p}_j(t) \rangle_{j\in\partial_i} \label{model_p}
\end{eqnarray}
where $\Pi$ normalizes vectors to unit length, ${\cal R_\eta}$ rotates vectors by a random angle distributed uniformly in a range $(-\pi\eta;\pi\eta]$, and the average is over the particles $j$ present in $\partial_i$,
the disk of radius $r_0 = 1$ centered on ${\bf r}_i$.
We checked that the results presented in the following are not sensitive to the value of $\alpha$, provided $0<\alpha<1$.
Therefore, in the following, we only use the numerically-convenient value $\alpha=\frac{1}{2}$.
With $\alpha$ fixed, the main parameters remain those of classic Vicsek-style models, the mean density of particles $\rho_0$, and the noise amplitude $\eta$.

\begin{figure}[t!]
	\includegraphics[scale=0.115]{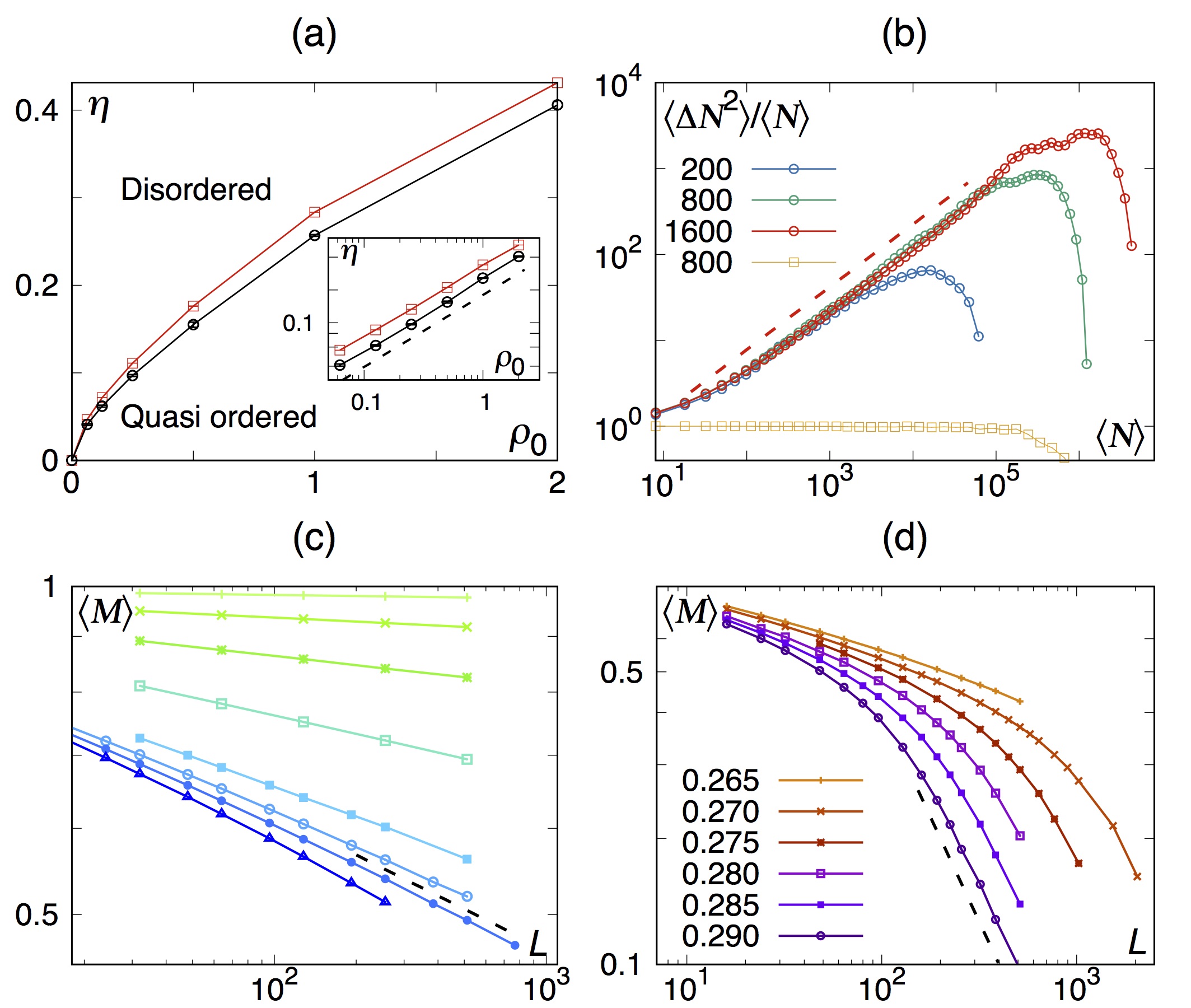}
	\caption{\small{(a): Phase diagram in the ($\rho_0,\eta$) plane. The asymptotic order-disorder transition line is shown in black. The red curve reports
the location of the susceptibility peak	 $\eta_{\chi}$ measured for $L=256$ (error bars are smaller than symbols). Inset: same data in logarithmic scales, with the dashed line marking slope $0.66$.
	(b):  Variance $\langle\Delta N^2\rangle$ over mean $\langle N\rangle$ of the number of particles present in sub-systems in the quasi-ordered (circles, $\eta = 0.2$) and disordered (squares, $\eta = 0.5$) phases for various system sizes $L$ ($\rho_0=2$). The dashed line corresponds to $\zeta=1.73$.
	(c) and (d): Averaged magnetization as a function of system size for several noise values in the quasi-ordered and disordered phases ($\rho_0=1$).
	In the quasi-ordered state, $\langle M\rangle$ decays algebraically, with an $\eta$-dependent exponent $\kappa(\eta)$.
	Curves in (c) correspond to, from top to bottom, $\eta = 0.05$, $0.1$, $0.15$, $0.2$, $0.24$, $0.25$, $0.255$ and $0.26$. In (c) (resp. (d)) the dashed black line marks slope $-\frac{1}{8}$ (resp. $-1$). }}
	\label{fig1}
\end{figure}

We numerically determined the phase diagram of our model in the $(\rho_0,\eta)$ plane (Fig.\ref{fig1}a). We find a single transition line from the disordered
gas observed at strong noise and/or low density to a phase with global ordering of polarities, characterized, at finite system size,
by a finite average value of the magnetization $M(t)=|\langle {\bf p}_i (t) \rangle_i|$.
Contrary to the other known classes mentioned in the introduction, we do not see any sign of phase separation. The transition
seems continuous, with only quasi-long-range order: the magnetization decreases algebraically with system size,
$\langle M \rangle_t \sim L^{-\kappa(\eta)}$ with $\kappa$ increasing continuously with $\eta$ (Fig.\ref{fig1}c).
At strong-enough noise, a crossover to a fully disordered phase characterized by $\langle M \rangle_t \sim 1/L$ is observed at large-enough sizes
(Fig.\ref{fig1}d). Like in all known orientationally-ordered dry active matter phases, giant number fluctuations are present (Fig.\ref{fig1}b):
the variance $\langle\Delta N^2\rangle$ of the number of particles in a sub-system containing on average $\langle N \rangle$ particles
scales faster than $\langle N \rangle$. We find that $\langle\Delta N^2\rangle \sim \langle N \rangle^{\zeta}$ with $\zeta = 1.73(3)$, a value similar to
those reported for the other classes \cite{chate2008collective,ginelli2010large,ngo2014large}.

\begin{figure}[t!]
	\includegraphics[scale=0.115]{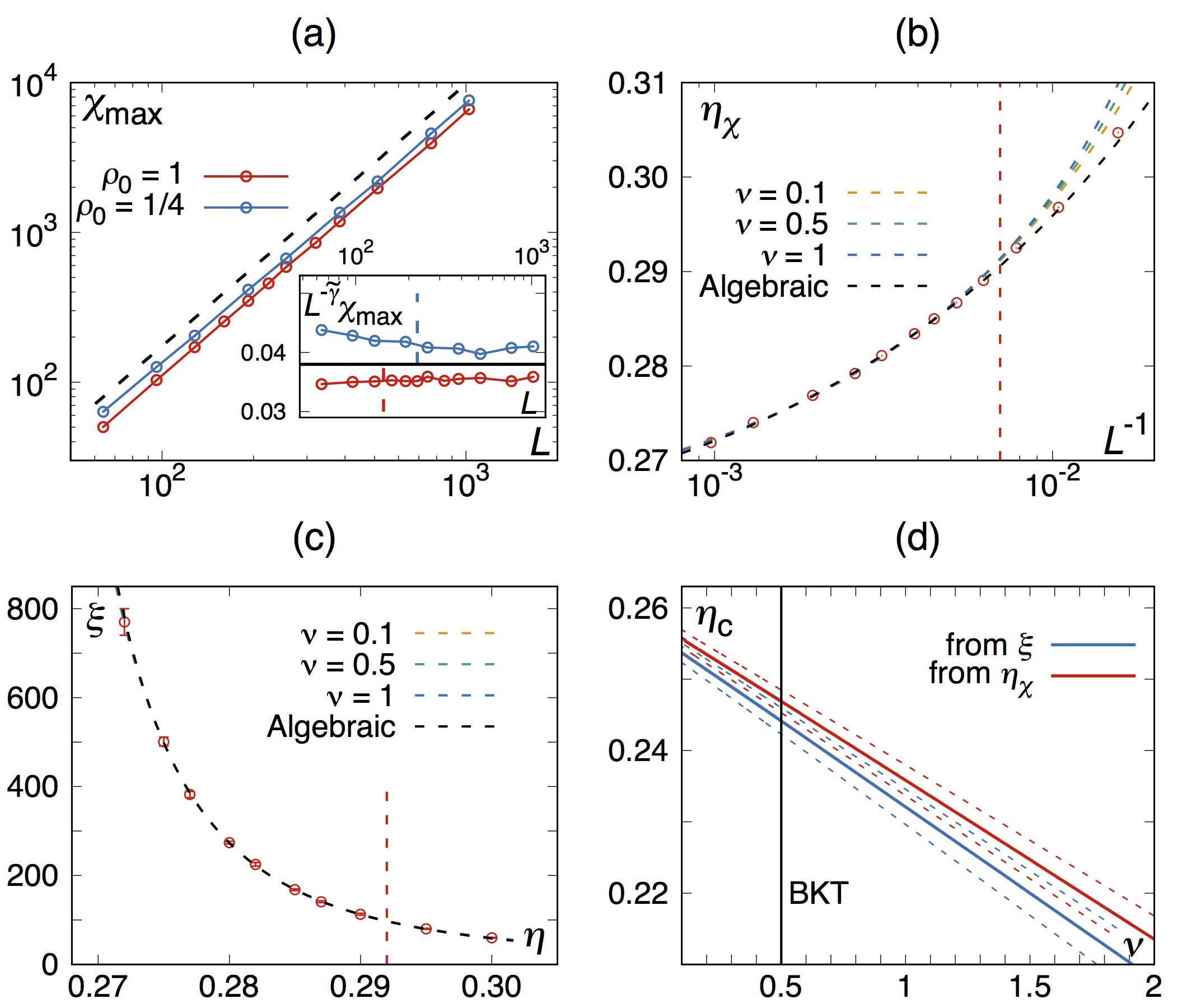}
	\caption{\small{(a): Susceptibility peak maximum $\chi_{\rm max}$ vs. system size $L$. The dashed line has slope $1.75$.
	Inset: same data scaled with $L^{-\tilde{\gamma}}$ with $\tilde{\gamma} = 1.75$. Here and in (b,c) the vertical dashed line delimits system sizes below which the scaling regime is not reached (these points are not used for the fits in (b), (c) and (d)).
	(b): Position of $\chi_{\rm max}$ vs. $L$. The different lines are fits using \eqref{chi_loc_KT} (KT-like scaling) for several values of $\nu$ and \eqref{Alg_fit} (algebraic scaling).
	(c): same as (b) but for divergence of the correlation length with noise.
	(d): asymptotic threshold $\eta_{\rm c}$ obtained from fits of $\xi$ and $\eta_\chi$ with BKT-like scaling \eqref{KT_fit} and \eqref{chi_loc_KT} varying the exponent $\nu$. The dashed lines represent the confidence intervals on $\eta_c$ given by the fits.
	}}
	\label{fig2}
\end{figure}

We now derive hydrodynamic equations for the Vicsek-shake class from our microscopic model.
Encouraged by its overall success in the other cases, we adopt the Boltzmann-Ginzburg-Landau approach \cite{bertin2006boltzmann,bertin2009hydrodynamic,peshkov2014boltzmann}.
We write two coupled Boltzmann equations for the single-body distributions $f_\pm({\bf x},\theta,t)$ of ``$+$" and ``$-$" particles, {\it i.e.} those which, respectively,
currently move along or against their polarity:
\begin{equation}
\partial_t f_\pm \pm {\bf e}(\theta)\!\cdot\! {\bm\nabla} f_\pm = a (f_\mp \!-\! f_\pm) + I_\text{dif}[f_\pm] + I_\text{col}[f_\pm,f_\mp]
\label{Boltzmann}
\end{equation}
where ${\bf e}(\theta)$ in the material derivatives of $f_\pm$ is the unit vector along $\theta$
\footnote{The speed $v_0$ has been set to unity without loss of generality.},
$a$ is the exchange rate between the two
subpopulations (akin to the microscopic reversal probability $\alpha$), and the self-diffusion and collisional integrals read:
\begin{equation}
I_\text{dif}[f_\pm] \!= \!-\! f_\pm \!+\!\! \int_0^{2\pi} \!\!\!\!\! d\theta'  \!\! \int_{-\infty}^{\infty} \!\!\!\!\! d\sigma  f_\pm(\theta')P_\eta(\sigma)\delta_{2\pi}(\theta\!-\!\theta'\!-\!\sigma)
\label{Idiff}
\end{equation}
\begin{eqnarray}
I_\text{col}[f_+,f_-] & = & \int_0^{2\pi} \!\!\! d\theta_1\int_0^{2\pi} \!\!\! d\theta_2 \int_{-\infty}^{\infty} \!\!\! d\sigma P_\eta(\sigma)f_+(\theta_1) \nonumber\\
&\!\!\!\!\!\!\!\!\!\!\!\!\!\!\times & \!\!\!\!\!\! \left[f_+(\theta_2)K^+(\theta_1\!-\!\theta_2) + f_-(\theta_2)K^-(\theta_1\!-\!\theta_2)\right] \nonumber\\
&\!\!\!\!\!\!\!\!\!\!\!\!\!\!\times & \!\!\!\!\!\! \left[\delta_{2\pi}(\theta-\Psi(\theta_1,\theta_2)-\sigma) - \delta_{2\pi}(\theta-\theta_1) \right]
\label{Icoll}
\end{eqnarray}
where $P_\eta(\sigma)$ is the noise distribution of variance $\eta^2$,
$\delta_{2\pi}$ is the Dirac comb distribution of period $2\pi$,
and $\Psi(\theta_1,\theta_2) = \text{Arg}({\bf e}(\theta_1)+{\bf e}(\theta_2))$ is the ferromagnetic alignment rule of polarities.
The kernels $K^\pm(\theta_1-\theta_2) = |{\bf e}(\theta_1)\mp{\bf e}(\theta_2)|$ are different and are used depending on whether the two colliding particles belong to the same population or not.

Hydrodynamic equations are derived from \eqref{Boltzmann} by expanding the distributions in angular Fourier modes: $f^\pm = \sum f_k^\pm \exp(-ik\theta)/(2\pi)$, and truncating and closing the resulting hierarchies in a controlled way.
In the classic Vicsek model the remaining hydrodynamic (or slow) fields correspond to the first two angular modes, {\it i.e.} density and polarity/velocity.
Here these fields are the zeroth and first modes of the sum $f=f_++f_-$, {\it i.e.} the density and polarity of the total population,
while the velocity field, now distinct from polarity, is the first mode of the difference $g=f_+-f_-$.
Rewriting the Boltzmann equations in terms of $f$ and $g$ modes, we obtain
\begin{eqnarray}
\partial_t f_k & + & \frac{1}{2}\left(\nabla^*g_{k+1} + \nabla g_{k-1}\right) = \left(P_k - 1\right)f_k \nonumber \\
\label{Fourier_f_Boltzmann}
 & & + \sum_{q=-\infty}^{+\infty} A_{k,q}f_qf_{k-q} + B_{k,q}g_qg_{k-q}\\
\partial_t g_k & + & \frac{1}{2}\left(\nabla^*f_{k+1} + \nabla f_{k-1}\right) = \left(P_k - 1 - 2a\right)g_k \nonumber \\
\label{Fourier_g_Boltzmann}
& & + \sum_{q=-\infty}^{+\infty} C_{k,q}f_qg_{k-q}
\end{eqnarray}
where the complex gradient $\nabla = \partial_x + i\partial_y$,
$P_k= \int d\sigma P_\eta(\sigma)\exp(ik\sigma)$,
and all other coefficients are listed in \cite{SUPP}.
Setting $f_1 \sim \varepsilon$ near onset of polar order, Eqs.~\eqref{Fourier_f_Boltzmann} and \eqref{Fourier_g_Boltzmann} impose the following
scaling ansatz \cite{peshkov2014boltzmann}:
\begin{eqnarray}
\delta\rho = \rho - \rho_0 \sim g_0 &\sim & \varepsilon \;  ;  \; |f_k| \sim |g_k|\sim\varepsilon^{k} \;\; \forall k>0\quad \label{scaling}\\
\nonumber \\
& & \partial_t \sim \nabla \sim \varepsilon
\end{eqnarray}
At the first non trivial order, $\varepsilon^3$, we get equations for $\rho$, $g_0$, $f_1$, $g_1$, $f_2$ and $g_2$.
The last two fields can then be enslaved to the four remaining ones, yielding:
\begin{eqnarray}
\label{H_density}
\partial_t\rho & = & -\Re\left(\nabla^* g_1\right)\\
\label{H_g0}
\partial_t g_0 & = & -2ag_0-\Re\left(\nabla^* f_1\right)
\end{eqnarray}

\begin{widetext}
\begin{eqnarray}
\label{H_f1}
\partial_t f_1 & = & (\mu_1[\rho] \!-\! \xi |f_1|^2 \!-\! \delta|g_1|^2) f_1
						+ \alpha\Delta f_1 + \left( \gamma[g_0] - \beta f_1^*g_1 \right)g_1 -\frac{1}{2}\nabla g_0
						+ \eta_1 f_1^*\nabla g_1 + \eta_2 g_1^*\nabla f_1 + \eta_3\nabla^*(f_1g_1)\\
\label{H_g1}
\partial_t g_1  & = & (\nu_1[\rho] \!-\! \tau|g_1|^2 \!-\! \omega|f_1|^2) g_1 + \lambda\Delta g_1
						 + \left(\kappa[g_0] - \chi g_1^*f_1\right)f_1 - \frac{1}{2}\nabla\rho
						 + \sigma_1 g_1^*\nabla g_1 + \sigma_2 f_1^*\nabla f_1+ \sigma_3 \nabla^*f_1^2 + \sigma_4\nabla^*g_1^2 \quad\quad
\end{eqnarray}
\end{widetext}
where all coefficients, expressed as functions of the microscopic parameters $\rho_0$, $\sigma$ and $a$ are listed in \cite{SUPP}
while their dependence on local density and $g_0$ has been made explicit.
Eqs.~\eqref{H_density} to \eqref{H_g1} can be seen as two coupled Toner-Tu equations \cite{toner1995long}
\footnote{The $g$ fields equations both have a negative linear coefficient regardless of density, noise and reversal rate values.
In the fast reversal limit these fields can therefore be enslaved to $\rho$ and $f_1$.
In \cite{SUPP} we derive equations directly in this limit and show that
they behave similarly to the full equations.}.
Note that density is {\it not} advected by the order field $f_1$, but by $g_1$, in strong contrast to the classic polar case.
Since $\nu_1[\rho]<0$ and $\mu_1[\rho]$ can change sign, the transition, as expected, is given by $\mu_1[\rho_0]=0$,
defining a line in the $(\rho_0,\eta)$ plane that goes to the origin as $\sqrt{\rho_0}$. Furthermore, since
$\mu_1[\rho]$ does not depend on $a$, this line is insensitive to the reversal rate, in agreement with the microscopic model.
When $\mu_1[\rho_0]<0$  the homogeneous disordered solution $\rho = \rho_0$, $f_1 = g_0 = g_1 = 0$ is linearly stable, and becomes
unstable when $\mu_1[\rho_0]>0$. It is then replaced by the homogeneous
ordered solution $\rho = \rho_0$, $g_0 = g_1 =0$, $f_1 = \sqrt{\mu_1[\rho_0]/\xi}$.
We studied its linear stability semi-numerically (see \cite{SUPP} for details) and analytically in the long wavelength limit (not shown).
Apart from a pocket of weak, spurious instability at small $a$ and low noises, it is essentially stable as soon as $\mu_1[\rho_0]>0$
\footnote{A small region of weak instability is present deep in the
  ordered phase for small reversal rate $a$.
  It quickly disappears upon increasing $a$. Furthermore we found
  that no such residual instability exists at the kinetic level, i.e. when
  considering (\ref {Fourier_f_Boltzmann}) and (\ref {Fourier_g_Boltzmann})
  with a large number of modes. We conclude that it is spurious (see \cite{SUPP} for details).}.

\begin{figure*}[t!]
	\includegraphics[scale=0.22]{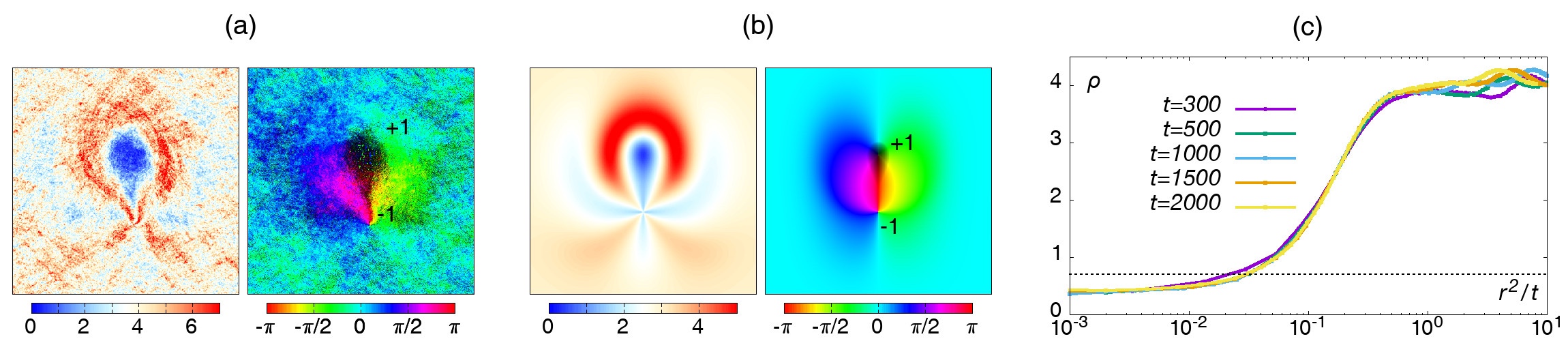}
	\caption{\small{(a) and (b):  Annihilation of an initially prepared pair of defects deep in the ordered phase.
	(a) Snapshots of density and polarity orientation fields from our microscopic model ($\rho_0=4$, $\eta=0.2$, $L^2=1024^2$ and $t=40000$).
	(b) Same as (a) but from a simulation of hydrodynamic equations (\ref{H_density}$-$\ref{H_g1}) ($\rho_0=3$, $\eta=0.4$, $a = 1$, $L^2=256^2$, $t =2500$).
	(c): Time-rescaled radial density profiles around the $+1$ defect in microscopic simulations (same parameters as in (a)).
	The dashed line indicates the transitional density $\rho_0^c(\eta)$ corresponding to $\eta=0.2$.}}
	\label{fig3}
\end{figure*}

The analysis above confirms, at the mean-field level, the absence of the generic instability leading to
the phase separation scenario in other classes of dry aligning active matter.
Here we have a single transition line separating polar order from the disordered phase.
Order (field $f_1$) and density are advected by the auxiliary field $g_1$, and thus the mechanism proven by Toner and Tu to be responsible
for the possibility of true long-range order is absent \cite{toner1995long,toner1998flocks}. With fluctuations, polar order is only quasi-long range, as in equilibrium.
Our problem thus possesses many of the hallmarks of the $XY$ model.
We now investigate whether this extends to the nature of the transition, {\it i.e.} whether
it is in the well-known Berezinskii-Kosterlitz-Thouless (BKT) universality class.
The BKT transition
is characterized by an essential divergence of the correlation length $\xi$ when approaching the critical point $\eta_{\rm c}$
from the disorder side, together with the scaling of the susceptibility $\chi$ with $\xi$ \cite{kosterlitz1973ordering,kosterlitz1974critical,gupta1988phase}
\begin{equation}
\log \xi \!\sim\! (\eta \!-\! \eta_{\rm c})^{-\nu} \, ; \; \chi \!\sim\! \xi^{\tilde{\gamma}}  \,\;{\rm with}\;
\chi \!=\! L^2(\langle M^2 \rangle \!-\! \langle M \rangle^2)    \label{KT_fit}
\end{equation}
with $\nu = \frac{1}{2}$ and $\tilde{\gamma}=\gamma/\nu=\frac{7}{4}$.
At finite size $L$, $\chi$ exhibits a maximum $\chi_\text{max}(L)$ located at $\eta_\chi(L)$.
Increasing $L$, $\chi_{\rm max}$ diverges, and $\eta_\chi$ converges to $\eta_{\rm c}$ like:
\begin{equation}
\chi_{\rm max}(L) \sim L^{\tilde{\gamma}} \;;\;\;
\eta_\chi(L) - \eta_{\rm c} \sim (\log(L)-a)^{-1/\nu} \;. \label{chi_loc_KT}
\end{equation}

We have measured the dependence of $\langle M \rangle$ and $\chi$ on $\eta$ for various systems sizes, all this at various global density values, but focussing
most of our numerical effort on $\rho_0=1$.
As shown in Fig.~\ref{fig2}a, the susceptibility peak $\chi_{\rm max}$ does diverge algebraically with an exponent $\tilde{\gamma}=1.755(6)$
in full agreement with the BKT/Ising value $\frac{7}{4}$.
The peak location $\eta_\chi$ is reasonably well fitted by Eq.~\eqref{chi_loc_KT} with $\nu=\frac{1}{2}$, yielding an estimate of the asymptotic threshold
$\eta_{\rm c} = 0.247(2)$ (Fig.~\ref{fig2}b).

We defined the correlation length $\xi$ as the crossover scale marking the beginning of the $1/L$ decay in $\langle M \rangle (L)$ curves at fixed $\eta$
(see \cite{SUPP} for details).
The divergence of $\xi$ with decreasing $\eta$ is well fitted by Eq.~\eqref{KT_fit} with $\nu=\frac{1}{2}$ (Fig.~\ref{fig2}c).
This fit yields an estimate $\eta_{\rm c} = 0.244(2)$ (barely) compatible with that obtained from the susceptibility.
However, for $0.244<\eta<0.247$, $\langle M \rangle$ decreases with exponent $\kappa(\eta) \simeq 0.100(5)$ {\it incompatible} with the BKT value $\frac{1}{8}$
(see Fig.~\ref{fig1}c).
 Repeating the procedure for $\rho_0 = \frac{1}{2}$ and $2$ we reach the same conclusion and find {\it different} values of $\kappa(\eta_{\rm c})$, respectively $0.089(6)$ and $0.117(2)$.

Allowing now $\nu$ to vary in a range [$0.1,2$], we find fits of the variations of $\eta_\chi$ and $\xi$ as convincing as for the BKT value $\nu=\frac{1}{2}$.
Interestingly, the two independent estimates of $\eta_{\rm c}$ then become closer to each other as $\nu\to0$ (Fig.~\ref{fig2}d).
This suggests an {\it algebraic} divergence for $\xi$ at threshold. We therefore redefine the $\nu$ exponent as that of a standard second-order
phase transition:
\begin{equation}
\xi \sim (\eta-\eta_{\rm c})^{-\nu} \; ; \;\; \eta_\chi (L) - \eta_{\rm c} \sim  L^{-1/\nu} \label{Alg_fit}
\end{equation}
Fitting our data accordingly, we obtain better fits for both $\xi$ and $\eta_\chi$ and, importantly, fully-compatible threshold values at which, moreover,
$\kappa(\eta_c)\simeq\frac{1}{8}$.
Imposing a common value for the asymptotic threshold, both datasets give the same estimate of $\nu$,
and we finally conclude that $\eta_{\rm c}=0.257(1)$ with $\nu=2.4(1)$.
From these values, we compute $\beta/\nu$ using a collapse of the magnetization curves (shown in \cite{SUPP}) and find a value
fully compatible with $\beta/\nu = \frac{1}{8}$, which satisfies the hyperscaling relation $2\beta/\nu + \gamma/\nu = d$ with $d=2$.
Using data obtained at various global densities, we find the same estimates of $\gamma/\nu$ and $\beta/\nu$,
although our estimate of $\nu$ shows some variation due to its sensitivity to the estimated value of $\eta_{\rm c}$.
The asymptotic threshold values thus obtained behave as $\eta_{\rm c}\sim\rho_0^h$ with $h\sim 0.66$, in clear departure
from the mean-field value $\frac{1}{2}$ (Fig.~\ref{fig1}a).

Our numerical analysis leads us to conclude that the transition to polar order exhibited by our system is {\it not} of the BKT type.
In the $XY$ model, this transition is closely related to the (effective) Coulomb interaction between
topological singularities that unbind and proliferate above the critical temperature \cite{kosterlitz1973ordering,kosterlitz1974critical}.
Detecting these defects in simulations of our model is made very difficult, if not impossible,
by the presence of strong density fluctuations. Indeed, the very existence of topologically constrained defects
requires that order can be defined everywhere. Here, the local order is hard to measure in sparse regions,
and even impossible to define if the local density is below the ordering threshold
$\rho_0^{\rm c}(\eta)$, the transitional density found by varying $\rho_0$ keeping $\eta$ fixed.
One can nevertheless study the fate of defects from carefully prepared initial configurations containing a $\pm1$ pair.
Running the model deep in the ordered phase, we observe that the positive defect expels particles from its core
and is quickly transformed into a sparse, almost void region whose diameter grows like $\sqrt{t}$ (Fig.~\ref{fig3}c).
After some time, this region has become sufficiently large so that it reaches the negative defect
and the system eventually repairs itself (Fig.~\ref{fig3}a, movie in \cite{SUPP}).
This mean-field behavior is also observed in simulations of the (deterministic) hydrodynamic equations (Fig.~\ref{fig3}b, movie in \cite{SUPP}).
It is intrinsically related to the coupling between density and order, and indicates that the very idea of topologically-bound point defects is not
relevant in our system. Closer to the transition, we of course expect fluctuations to play a major role, but this conclusion should still hold.
This is the topic of ongoing work.

To summarize, we have shown that the collective behavior of active particles with velocity reversals that align ferromagnetically their polarities
is different from that of other classes of dry, dilute, aligning active matter. This new class is characterized by the emergence of a
phase with quasi-long-range polar order and anomalous number fluctuations. Like in the $XY$ model, scaling exponents vary continuously
in this phase, but the transition point to order shows algebraic divergences governed by $\nu=2.4(1)$, not the essential singularity of the BKT class.
Nevertheless, the exponent ratios $\beta/\nu$ and $\gamma/\nu$ take the BKT/Ising values $\frac{1}{8}$ and $\frac{7}{4}$.
These results constitute the first case where the phase separation
scenario at play in most dry aligning active matter systems is prevented ``structurally", not by imposing incompressibility \cite{chen2015critical} or by resorting
to non-metric neighbors \cite{ginelli2010relevance,peshkov2012continuous}. The estimated exponents seem to correspond to a new type of non-equilibrium critical point.

We thank Francesco Ginelli, Cesare Nardini, and Julien Tailleur for helpful discussions and a critical reading of our manuscript.
This work is partially supported by ANR project Bactterns, FRM project Neisseria, and
the National Natural Science Foundation of China (No. 11635002 to X.S and H.C.; No. 11474210 and No. 11674236 to X.S.).

\bibliography{Biblio}
\bibliographystyle{prsty}

\end{document}